# Magnetic spin imaging under ambient conditions with sub-cellular resolution

S. Steinert[1], F. Ziem[1], L. Hall[2], A. Zappe[1], M. Schweikert[3], A. Aird[1], G. Balasubramanian[4],

L. Hollenberg[2], J. Wrachtrup[1*]

1. 3rd Institute of Physics and Research Center SCOPE, University Stuttgart, Pfaffenwaldring 57, 70569 Stuttgart, Germany
2. Centre for Quantum Computation and Communication Technology, School of Physics, University of Melbourne, Parkville, Victoria 3010, Australia
3. Biologisches Institut, University Stuttgart, Pfaffenwaldring 57, 70569 Stuttgart, Germany
4. Nanoscale Spin Imaging Group, Max-Planck Institute for Biophysical Chemistry, Am Fassberg 11, 37077, Göttingen, Germany.

*To whom correspondence should be addressed. Email: j.wrachtrup@physik.uni-stuttgart.de, s.steinert@physik.uni-stuttgart.de,

**Abstract:**

Measuring spins is the corner stone of a variety of analytical techniques including modern magnetic resonance imaging (MRI). The full potential of spin imaging and sensing across length scales is hindered by the achievable signal-to-noise in inductive detection schemes. Here we show that a proximal Nitrogen-Vacancy (NV) ensemble serves as a precision sensing array. Monitoring its quantum relaxation enables sensing of freely diffusing and unperturbed magnetic ions in a microfluidic device. Multiplexed CCD acquisition and an optimized detection scheme enable direct spin noise imaging under ambient conditions with experimental sensitivities down to 1000 statistically polarized spins, of which only 35 ions contribute to a net magnetization, and 20 s acquisition time. We also demonstrate imaging of spin labeled cellular structures with spatial resolutions below 500 nm. Our study marks a major step towards sub-µm imaging magnetometry and applications in microanalytics, material and life sciences.

**Main text**

Sensing weak magnetic fields is a key technology, particularly in biomedical and chemical science where the detection of Boltzmann polarized spin ensembles has proven its versatility in MRI. However, high resolution magnetic imaging is mainly hindered by inductive detection requiring a sufficient magnetic flux from the sample to be sensed by the pickup coil. Increasing spatial resolution reduces the detection volume until the net magnetization is exceeded by statistical polarization [1]. The latter scales with $\mu\sqrt{n}/n$, where $n$ is the number of spins and $\mu$ their magnetic moment. The resulting average field $\langle B \rangle \approx 0$ and its stochastic nature render common detection schemes ineffective. Precision magnetometers such as super-conducting-interference devices [2], magnetic-resonance-force-microscopy [3], atomic vapors [4] and remote detection schemes [5] provide exceptional sensitivities reaching a few fT/√Hz. Their relatively large size and/or operational conditions prohibit sensing of small numbers of spin magnetic moments at ambient conditions and sub-μm resolution and thus have hampered widespread applications so far.

Sensors based on Nitrogen-Vacancy (NV) centers in diamond are a possible alternative due to its atomic size. This allows placement of the sensor in few nm proximity to the sample while retaining superior volume-to-sensitivity scaling, room temperature operation and optical readout [6–8]. NV sensing of external controlled DC and AC fields has subsequently been demonstrated [9–12]. Recent NV metrology demonstrations have shown the detection of spins within the diamond lattice [13] and surface spins [14]. However, NV metrology yet lacks the detection of spins present outside the diamond lattice.

Here we demonstrate sensing and imaging of stochastic magnetic fluctuations originating from freely diffusing electron spins such as paramagnetic oxygen ($O_2$, S=1), $MnCl_2$ (S=5/2) and Gadolinium ions ($Gd^{3+}$, S=7/2) in liquids, immobilized in polymers and linked specifically to cellular structures. Particularly the large magnetic moment of $Gd^{3+}$ has made the $Gd^{3+}$ chelate a prime candidate as relaxation contrast agent in MRI [15]. In the limit of low external magnetic fields ($0<B_0<10$ mT), the stochastic process of spin fluctuations of freely diffusing ions exhibit a zero-mean field, but measurable magnetic field fluctuations arise due to the statistical spin polarization causing a non-zero RMS field $\sqrt{\langle B^2 \rangle} > 0$ with random phase. Such random fields are difficult to detect, but the NV offers a striking technique for such random fields by monitoring its quantum relaxation [16].

As illustrated in Fig. 1, we employ an array of atomic sized NV sensors (~1000 μm$^{-2}$) with a calculated mean depth of $h$=6.7 nm [17]. The NV array is optically spin polarized >95 % [18] via the singlet state into the $|0\rangle$ ground state by a green laser pulse. After preparation of a distinct NV spin state, interaction with the environment leads to relaxation with rate $\Gamma = \Gamma_{int} + \Gamma_{env}$. $\Gamma_{int}$ includes diamond internal contributions due to spin impurities and lattice dynamics and $\Gamma_{env}$ those to which the system can deliberately be exposed. Relaxation may occur via the transverse and longitudinal relaxation channels characterized by their respective decay times $T_2$ and $T_1$. The transverse dephasing rate $\Gamma_2=1/T_2$ is increased by fluctuations at low frequencies (kHz to MHz) and is observed by spin echos of Hahn or CPMG type [19,20]. $\Gamma_1=1/T_1$ is susceptible to frequencies at the NV Larmor precession $\omega_0 = D \pm \gamma B_0$, where $\gamma$ is the gyromagnetic ratio and $B_0$ is a magnetic offset field. Due to the large NV zero-field splitting of $D$=2.87 GHz this is generally in the GHz range. $\Gamma_1$ describes the decay of a spin initially polarized into $|0\rangle$ to a thermal equilibrium mixture. Both rates can be read out optically as shelving of the $|\pm1\rangle$ levels in the metastable singlet states is associated with a decrease in NV fluorescence intensity. The laser excitation is set to wide-field illumination and the red-shifted fluorescence response of the NV sensor is projected onto a CCD camera with a field of view of 60x60 µm². A microfluidic device provides experimental control over the aqueous $Gd^{3+}$ solution.

First, we demonstrate magnetic sensing of various chemical environments by probing both relaxation channels along one preferred NV axis ($B_0$=5 mT) in the presence of air, water and a solution of 1 M $Gd^{3+}$ (see Table 1). Evidently, $\Gamma_2$ probed by a Hahn echo shows some marginal changes in $T_2$ for $Gd^{3+}$, while multi-pulse CPMG81 is mildly responsive with a change in $T_2$ of 13.5 %. The longitudinal $T_1$, however, exhibits a prominent reduction of 94 % in the presence of $Gd^{3+}$ and was similarly responsive to dissolved $MnCl_2$ and with $O_2$ saturated water (Fig. S1). Focusing on the dominant $Gd^{3+}$, the relaxation rate $\Gamma_{Gd}$ induced by freely diffusing $Gd^{3+}$ depends largely on the RMS magnetic field and its spectral density $S_{Gd}(\omega)$. For a Gaussian process, the latter is given by $S_{Gd}(\omega) = \sqrt{2/\pi} \cdot f_{Gd}/[(\omega - \omega_0)^2 + f_{Gd}^2]$, where $\omega_0$ is the Larmor frequency of $Gd^{3+}$ and $f_{Gd}$ are the various fluctuation sources of the ions described in the following. Boltzmann polarization and $\omega_0$ can be neglected for the low $B_0$ fields applied. $S_{Gd}(\omega)$ is instead dominated by statistical polarization and substantial broadening effects of zero-mean fluctuations $f_{Gd}$ which is decomposed into $f_{Gd} = f_{dipole} + f_{vib} + f_{trans} + f_{rot}$, where $f_{dipole} = c_{Gd} \cdot 77$ GHz M$^{-1}$ denotes the concentration dependent dipole coupling strength between $Gd^{3+}$ (see SOM) and $c_{Gd}$ is the spin concentration in mol/l. Intrinsic vibrational spin relaxation of the complexed $Gd^{3+}$ ion yields a constant fluctuation of $f_{vib}$~50 GHz [21], while rotational motion $f_{rot}$ and translational diffusion $f_{trans}$ cause fluctuations of ~140 MHz. The combined dynamics broaden the spectral density causing $S_{Gd}(\omega)$ to be constant up to a few tens of GHz (Fig. 2B). Evidently, $S_{Gd}(\omega)$ is dominated by the fast intrinsic $Gd^{3+}$ relaxation $f_{vib}$ for $c_{Gd}$<0.5 M and is even broader at higher $c_{Gd}$ due to the elevated $Gd^{3+}$ coupling $f_{dipol}$. Resonant $Gd^{3+}$ induced fluctuations $B_x$ and $B_y$ (NV axis defines z) cause $\Gamma_1$ relaxation of the NV. Taking the trace over a purely mixed state of the dipolar coupled NV and $Gd^{3+}$ results in a magnetic field of

$$\langle B_{Gd}^2 \rangle = \langle B_x^2 \rangle + \langle B_y^2 \rangle = \frac{21 \cdot 10^3 \pi N_A c_{Gd}}{16 h^3} \left(\frac{\mu_0}{4\pi} \gamma_{NV} \gamma_{Gd}\right),$$

where $N_A$ is the Avogadro constant, $\mu_0$ the vacuum permeability, $h$ the mean depth of the NV centers and $\gamma_{NV} \approx \gamma_{Gd}$ (Fig. S4). The probability of finding the NV in the $|0\rangle$ state reads as

$$P_0(\tau) = \frac{1}{6}\left(2 + e^{-\Gamma_1^- \tau} + e^{-\Gamma_1^+ \tau} + 2e^{-(\Gamma_1^- + \Gamma_1^+)\tau}\right),$$

where $\Gamma_1^\pm$ are the individual decay rates of each sensitivity window $F_1^\pm$ (Fig. 2B) integrated over $S_{Gd}(\omega)$. The overall longitudinal decay rate $\Gamma_{1,Gd}$ of the longitudinal NV magnetization is then given by

$$\Gamma_{1,Gd} = -\frac{d}{d\tau}\bigg|_{\tau=0} P_0(\tau) \approx 2 \frac{f_{Gd} \langle B_{Gd}^2 \rangle}{f_{Gd}^2 + D^2}.$$

This theoretical prediction was experimentally verified by varying $c_{Gd}$ (Fig. 2C), showing excellent agreement with theoretical expectations for the non-trivial dependence on $c_{Gd}$. On nanometer length scales, the sample-sensor distance is a key parameter as $\Gamma_{1,Gd} \propto \langle B_{Gd}^2 \rangle \propto h^{-3}$ and thus represents a viable tool to access the average implantation depth of nitrogen ions (Fig. 2C). The relaxometric SNR=$\Gamma_{Gd}/\Gamma_{int}$ of $T_1$ is much higher as $\Gamma_{2,int} \gg \Gamma_{1,int}$ due to the low phonon density in diamond [22]. Hence, $T_1$ relaxometry is two orders of magnitude more sensitive in this context (Fig. S3) and employed from here on.

Next, we optimized the microfluidic detection by converting the $T_1$ signal directly into a measureable fluorescence. Instead of detecting the full relaxation curve, the fluorescent NV response at a single-$\tau$ point is utilized as direct concentration readout. Although maximal fluorescence contrast is reached at $\tau_{opt}$ (Fig. 2A), maximum sensitivity is achieved at $\tau \sim T_{1,Gd}/2$ for $c_{Gd}$<10 mM and approaches $T_{1,Gd}$ for higher $c_{Gd}$ (Fig. S8). Therefore, varying $Gd^{3+}$ levels can be monitored dynamically with temporal resolutions in the order of a second (Fig. 2D). For maximal sensitivity all four crystallographic NV orientations can be employed, since $\langle B_{Gd}^2 \rangle$ exhibits a random magnetic field orientation. This is achieved by measuring in near-zero field (50 µT earth field) and probing all NV orientations at $D$=2.87 GHz (Fig. S9), which effectively improves the fluorescent single-$\tau$ contrast and increases the number of sensing NV spins by a factor of four. We applied a statistical t-test to the fluorescence traces to identify the lowest concentration $c_{Gd}^{min}$ still significantly different to pure water. After $t_m$=20 s and an optimized $\tau$=400 µs, $c_{Gd}^{min}$ =250 µM and 80 µM for single and all four NV axes, respectively. The latter concentration corresponds to a $Gd^{3+}$ concentration of 1.4 ppm or $\langle B_{rms} \rangle = 1.5$ µT. In addition, the multiplexed CCD detection allows varying the effective detection voxel by changing the pixel binning. Although single pixel analysis can be done, the lower limit of the spatial resolution is given by the diffraction limit of light which corresponds to 430 nm or 4x4 pixels (Fig. S10). For such a small voxel, $c_{Gd}^{min} = 500$ µM. For fixed $t_m$ and CCD detection, $c_{Gd}^{min}$ depends on the shot-noise of detected photons and is thus improved by increasing the pixel binning. To determine the sensitivity in terms of number of spins per detection voxel via $n_{Gd} = c_{Gd}^{min} \cdot 10^3 \cdot N_A \cdot \Delta r_{xy}^2 \cdot \Delta r_z$, it is important to emphasize that $\Gamma_{1,Gd}$ integrates the signal from the spins above the surface while proximal ions ($r=h$) contribute most due to the $h^{-3}$ scaling. The experimental height dependency employing polymer spacers between NV and $Gd^{3+}$ yielded a decay of $\Gamma_{1,Gd}(r)$ to 1/e at $\Delta r_z$=15 nm (Fig. S11). Taking the experimental sensitivity to $c_{Gd}^{min}$=500 µM with a spatial resolution of $\Delta r_{xy}$=460 nm (4x4 pixels), this corresponds to an experimental detection of 1000 spins within $t_m$=20 s.

In the following, this unprecedented spatial-temporal resolution is used to image magnetic fluctuations originating from samples on the sensor array (Fig. 3A). A periodic grid of lithographically patterned $Gd^{3+}$ is easily resolvable in the reconstructed $T_1$ image (Fig. 3B). However, $t_m$ can be reduced by orders of magnitude applying the optimized single-τ detection still yielding higher contrast of magnetic imaging (Fig. 3C). For high-resolution magnetic imaging in biological samples, we have specifically labeled the plasma membrane of HeLa cells with caged $Gd^{3+}$ ions and an Alexa532 fluorophore via biotinylated poly-L-lysine. Since the NV sensor is most sensitive to proximal spins, we placed 150 nm thin ultramicrotome sections of labeled cells onto the diamond sensor. The control fluorophore indicates a successful label of the magnetic marker to HeLa cells (Fig. 3D), while the boundary of the cell is clearly present in the magnetic image (Fig. 3E). Furthermore, a line scan through the magnetically labeled plasma membrane verifies a spatial resolution of 430 nm corresponding to the inherent diffraction limit of the conventional widefield technique. A complete simulation of the single-τ detection based on shot-noise limited photon detection agrees with the experimental sensitivities (Fig. S8). The key advantage of magnetic spin labels is the potential for chemically selective spin contrast imaging as each spin label is expected to have a distinct *S(ω)*, e.g. by binding to a certain biological complex. Using the NV, the spectral density can be measured with in principle unlimited bandwidth as $\Gamma_1$ provides a narrow and via $B_0$ tunable sensitivity window ($F_1$ in Fig. 2B and experimental realization in Fig. S5).

The results presented here demonstrate high sensitive non-invasive spin sensing and imaging of unperturbed $Gd^{3+}$ ions at room temperature. Although other sensitive techniques have probed statistical polarization [3,23–25], we have shown the NV based relaxometric technique offers several advantages as it is operable under ambient conditions with no requirements on strong magnetic fields or ion manipulating radiofrequency pulses. The optically readable solid state sensor offers large-scale sensor technology and could be easily integrated into microfluidics facilitating lab-on-chip devices. Quantum relaxation of the NV sensor has the potential to emerge as a novel technique of high throughput analytical sciences and contrast enhanced optical-MRI reaching sub-μm length scales. A considerable improvement in sensitivity is possible by decreasing the NV depth *h* and the voxel size $\Delta r_{xy}$ equivalent to enhancing the spatial resolution (Fig. S8). With *h*=2.5 nm and typical resolutions of $\Delta r_{xy}$=50 nm for structured widefield illumination (SIM) [26], a sensitivity in the order of ten $Gd^{3+}$ spins is expected. Although requiring time consuming point-scanning, stimulated emission depletion (STED) would boost the $\Delta r_{xy}$ down to 8 nm [27] reaching even single spin sensitivities. The high temporal resolution of widefield magnetometry also favors sub-cellular visualization of label-free dynamic processes, for instance the production of free radicals in cell death, the regulation of homeostasis through ion channels or hemoglobin trafficking by imaging of paramagnetic oxygen.

**References and Notes:**

1. Bloch, F. Nuclear Induction. *Phys. Rev.* **70**, 460–474 (1946).
2. Clarke, J. & Braginski, A. I. *The SQUID Handbook: Applications of SQUIDs and SQUID systems*. (Wiley-VCH: 2006).
3. Rugar, D., Budakian, R., Mamin, H. J. & Chui, B. W. Single spin detection by magnetic resonance force microscopy. *Nature* **430**, 329–332 (2004).
4. Budker, D. & Romalis, M. Optical magnetometry. *Nat Phys* **3**, 227–234 (2007).
5. Bajaj, V. S., Paulsen, J., Harel, E. & Pines, A. Zooming In on Microscopic Flow by Remotely Detected MRI. *Science* **330**, 1078–1081 (2010).
6. Gruber, A. *et al.* Scanning Confocal Optical Microscopy and Magnetic Resonance on Single Defect Centers. *Science* **276**, 2012–2014 (1997).
7. Chernobrod, B. M. & Berman, G. P. Spin microscope based on optically detected magnetic resonance. *Journal of Applied Physics* **97**, 014903 (2005).
8. Degen, C. Nanoscale magnetometry: Microscopy with single spins. *Nat Nano* **3**, 643–644 (2008).
9. Taylor, J. M. *et al.* High-sensitivity diamond magnetometer with nanoscale resolution. *Nat Phys* **4**, 810–816 (2008).
10. Maze, J. R. *et al.* Nanoscale magnetic sensing with an individual electronic spin in diamond. *Nature* **455**, 644–647 (2008).
11. Balasubramanian, G. *et al.* Ultralong spin coherence time in isotopically engineered diamond. *Nat Mater* **8**, 383–387 (2009).
12. Steinert, S. *et al.* High sensitivity magnetic imaging using an array of spins in diamond. *Review of Scientific Instruments* **81**, 043705 (2010).
13. Neumann, P. *et al.* Quantum register based on coupled electron spins in a room-temperature solid. *Nat Phys* **6**, 249–253 (2010).
14. Grotz, B. *et al.* Sensing external spins with nitrogen-vacancy diamond. *New Journal of Physics* **13**, 055004 (2011).
15. Chan & Wong, W. Small molecular gadolinium(III) complexes as MRI contrast agents for diagnostic imaging. *Coordination Chemistry Reviews* **251**, 2428–2451 (2007).
16. Cole, J. H. & Hollenberg, L. C. L. Scanning quantum decoherence microscopy. *Nanotechnology* **20**, 495401 (2009).
17. Ziegler, J. F. *The Stopping and Range of Ions in Solids*. (Pergamon.P.: 1985).
18. Waldherr, G. *et al.* Dark States of Single Nitrogen-Vacancy Centers in Diamond Unraveled by Single Shot NMR. *Phys. Rev. Lett.* **106**, 157601 (2011).
19. Du, J. *et al.* Preserving electron spin coherence in solids by optimal dynamical decoupling. *Nature* **461**, 1265–1268 (2009).

20. de Lange, G., Wang, Z. H., Riste, D., Dobrovitski, V. V. & Hanson, R. Universal Dynamical Decoupling of a Single Solid-State Spin from a Spin Bath. *Science* science.1192739 (2010).doi:10.1126/science.1192739
21. Kruk, D., Kowalewski, J. & Westlund, P.-O. Nuclear and electron spin relaxation in paramagnetic complexes in solution: Effects of the quantum nature of molecular vibrations. *The Journal of Chemical Physics* **121**, 2215 (2004).
22. Jarmola, A., Acosta, V. M., Jensen, K., Chemerisov, S. & Budker, D. Temperature- and Magnetic-Field-Dependent Longitudinal Spin Relaxation in Nitrogen-Vacancy Ensembles in Diamond. *Phys. Rev. Lett.* **108**, 197601 (2012).
23. Müller, N. & Jerschow, A. Nuclear Spin Noise Imaging. *PNAS* **103**, 6790–6792 (2006).
24. Crooker, S. A., Rickel, D. G., Balatsky, A. V. & Smith, D. L. Spectroscopy of spontaneous spin noise as a probe of spin dynamics and magnetic resonance. *Nature* **431**, 49–52 (2004).
25. Degen, C. L., Poggio, M., Mamin, H. J., Rettner, C. T. & Rugar, D. Nanoscale magnetic resonance imaging. *Proceedings of the National Academy of Sciences* **106**, 1313 –1317 (2009).
26. Gustafsson, M. G. L. Nonlinear structured-illumination microscopy: Wide-field fluorescence imaging with theoretically unlimited resolution. *Proceedings of the National Academy of Sciences of the United States of America* **102**, 13081 –13086 (2005).
27. Rittweger, E., Han, K. Y., Irvine, S. E., Eggeling, C. & Hell, S. W. STED microscopy reveals crystal colour centres with nanometric resolution. *Nat Photon* **3**, 144–147 (2009).

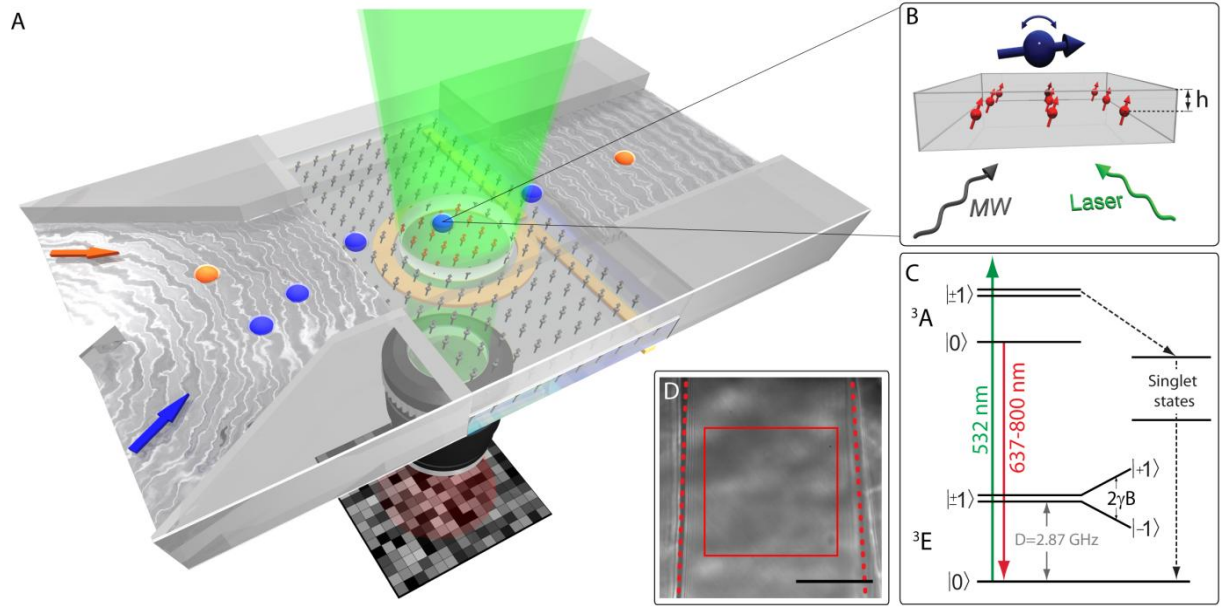

**Fig. 1. Widefield magnetometry with microfluidic detection. (A-B)** Widefield excitation (green) of NV spin ensemble (red arrows) and spatially resolved CCD detection of NV fluorescence. Homogeneous spin manipulation of the NV array is achieved via a lithographic Ω-structure (yellow). $Gd^{3+}$ ions (blue spheres in A-B) in aqueous solution are introduced using a microfluidic channel placed directly on top of the sensor proximal to the NV array with $h$=6.7 nm. **(C)** Energy level scheme of NV center illustrating the high fluorescent signal extending from 637-800 nm of the $|0\rangle$ sublevel. In contrast, $|\pm1\rangle$ states emit less photons due to a higher probability to enter the long-living singlet states. **(D)** Brightfield image of the microfluidic channel with channel boundaries (dotted lines) and 30x30 μm detection region of interest (ROI, red rectangle). Scale bar is 20 μm.

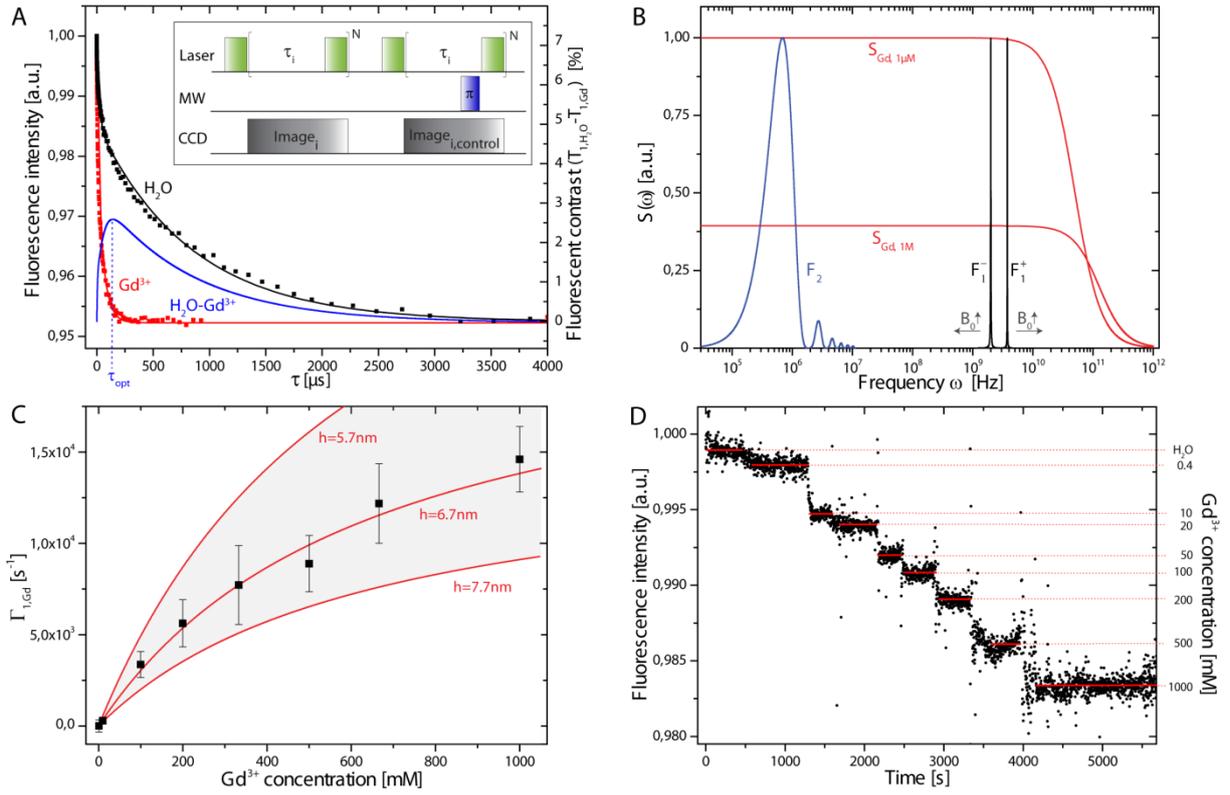

Fig. 2. Microfluidic $T_1$ relaxometry. (A) $T_1$ relaxation curves of the NV ensemble in the presence of water and 1 M $Gd^{3+}$ and the fluorescent contrast between both peaking at $\tau_{opt}$. The inset shows the measurement pulse sequence and normalization. (B) Spectral density $S_{Gd}(\omega)$ of $Gd^{3+}$ for two distinct concentrations (red curves) illustrating the broadening effect due to $Gd^{3+}$ coupling ($f_{dipole}$) at higher concentrations. While the sensitivity windows of $T_2$-decoherence ($F_2$, blue) is limited to low MHz fluctuations, $T_1$-relaxometry allows to probe a wide frequency range up to GHz with two sensitivity windows ($F_1^-$ and $F_1^+$, black). $F_1^\pm$ can be Zeeman-shifted via $B_0$ enabling experimental detection of $S_{Gd}(\omega)$. (C) Experimental relaxation rate $\Gamma_{1,Gd}$ in dependence on the $Gd^{3+}$ concentration and analytical predictions for specific mean implantation depths $h$ (red curves). Error bars represent standard deviation of three independent data sets. (D) Dynamic microfluidic single-$\tau$ detection ($\tau=100$ μs) of distinct $Gd^{3+}$ concentrations where a single data point is acquired within $t_m \sim 2$ s.

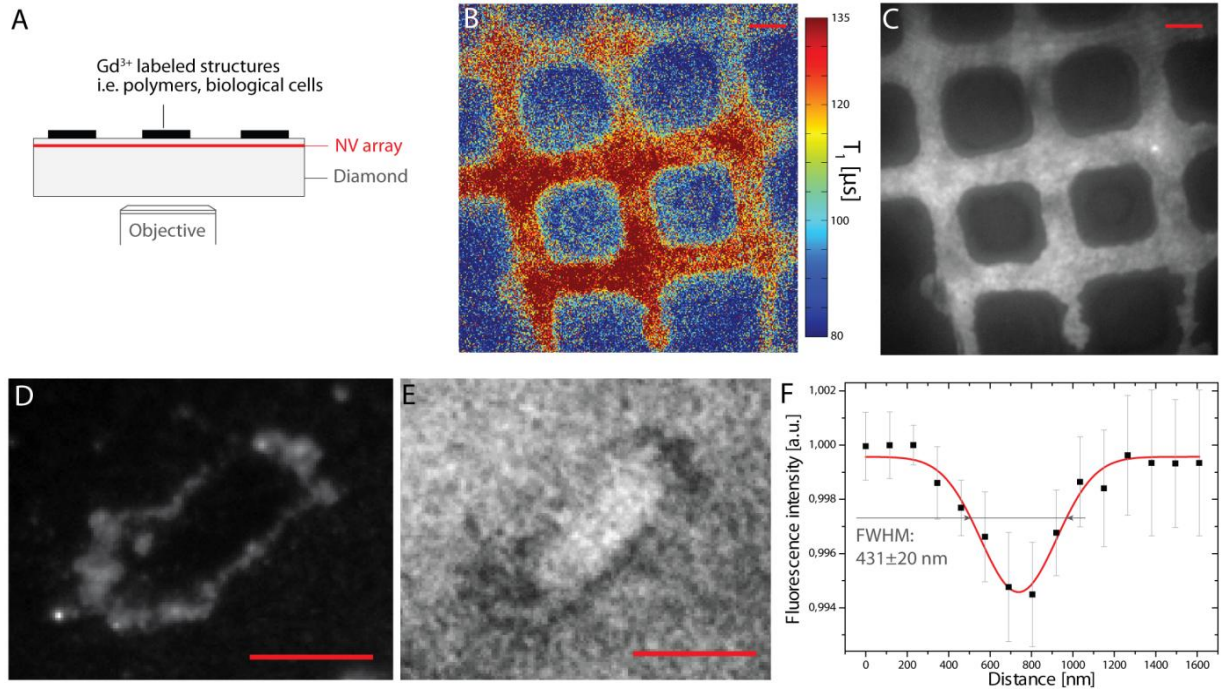

Fig. 3. Spin contrast imaging. (A) Schematic of magnetic spin imaging. (B) $T_1$ weighted image of lithographically patterned $Gd^{3+}$ grid (blue rectangular regions with low $T_1$) on top of diamond sensor. Three data sets each acquired within $t_m$=15 min were averaged and subsequently fitted to obtain $T_1$ for each pixel. (C) Single-τ imaging (τ=150 µs) directly yields dark areas where $Gd^{3+}$ is present due to the elevated NV relaxation. Although a single image ($t_m$=2 s) is sufficient to identify the pattern, the image shown was averaged for 10 min to enhance the contrast. (D) Fluorescent control image of an ultramicrotome cut (150 nm) of HeLa cells where the plasma membrane was labeled with Biotin-poly-L-Lysine-$Gd^{3+}$-DTPA-Alexa532. (E) Magnetic imaging via a single-τ measurement ($t_m$=15 min, τ=440 µs, $B_0$~50 µT) evidencing the presence of magnetic $Gd^{3+}$ at the cell membrane (dark structures). (F) Line scan through plasma membrane shown in (E) demonstrating spatial resolution of 431 nm. Error bars represent standard deviation of six independent line scans. Scale bars in B,C,D,E are 5 µm.

Table 1. Transverse ($T_2$) and longitudinal ($T_1$) NV spin relaxation in various environments (air, water and 1 M $Gd^{3+}$). The change ΔT in the corresponding relaxation times is given by ΔT=($T_{air}$-$T_{Gd}$)/$T_{air}$ , $\Gamma_{Gd}$=1/$T_{Gd}$-1/$T_{air}$ corresponds to the induced relaxation in the presence of 1 M $Gd^{3+}$ and $\Gamma_{int}$=1/$T_{air}$ is the diamond intrinsic relaxation. Since $\Gamma_{int}$ competes with the $Gd^{3+}$ induced relaxation, the relaxometric signal-to-noise ratio is given by SNR=$\Gamma_{Gd}$/$\Gamma_{int}$.

|  |  | $T_2$-Hahn | $T_2$-CPMG81 | $T_1$ |
|---|---|---|---|---|
| Air | [$\mu s$] | 1.92±0.04 | 42.9±1.8 | 1155±87 |
| dH2O | [$\mu s$] | 1.91±0.03 | 42.1±2.1 | 906±75 |
| $Gd^{3+}$ | [$\mu s$] | 1.85 ± 0.05 | 37.1±2 | 70±4 |
| ΔT | [%] | 3.7 | 13.5 | 94.0 |
| SNR= $\Gamma_{Gd}/\Gamma_{int}$ |  | 0.04 | 0.16 | 15.5 |